\documentclass[prl,aps,twocolumn,groupedaddress,showpacs,floatfix]{revtex4}
\usepackage{amsmath,amssymb,multirow,epsfig,bm}

\newcommand{\etal}{{\it et al.,\;}}
\newcommand{\beq}{\begin{equation}}
\newcommand{\eeq}{\end{equation}}
\newcommand{\bea}{\begin{eqnarray}}
\newcommand{\eea}{\end{eqnarray}}

\newcommand{\nn}{\nonumber}
\newcommand{\benn}{\begin{displaymath}}
\newcommand{\eenn}{\end{displaymath}}

\begin{document}

\title{Large Amplitude Dynamics of the Pairing Correlations in a Unitary Fermi Gas} 
 
\author{ Aurel Bulgac and Sukjin Yoon}
\affiliation{Department of Physics, University of Washington, Seattle, WA 98195--1560, USA}

\begin{abstract}
A unitary Fermi gas has a surprisingly rich spectrum of large amplitude modes of the pairing field alone, which defies a description within a formalism involving only a reduced set of degrees of freedom, such as quantum hydrodynamics or a Landau-Ginzburg-like description.  These modes are very slow, with oscillation frequencies well below the pairing gap, which makes their damping through quasiparticle excitations quite ineffective. In atomic traps these modes couple naturally with the density oscillations,  and the corresponding oscillations of the atomic cloud are an example of a new type of collective mode in superfluid Fermi systems. They have lower frequencies than the compressional collective hydrodynamic oscillations, have a non-spherical momentum distribution, and could be excited by a quick time variation of the scattering length. 
\end{abstract}

\date{\today}


\maketitle

While it is natural to expect the presence of hydrodynamic collective modes in a unitary gas \cite{ketterle}, the existence of collective oscillations of large amplitude nonlinear modes comes largely as a surprise. One reason is that these soliton-like modes do not emerge from a simplified quantum hydrodynamic or Landau-Ginzburg-like description of these systems. 
The quest for reducing the quantum description of a complex many-body system of particles to a relatively small number of degrees of freedom is one of the ongoing efforts in all physics subfields. In chemistry one would like to have an accurate description of complex molecules with many atoms and many more electrons in terms of a few wisely chosen relevant degrees of freedom (rotations, vibrations, bond stretching, shape, etc.).  In nuclear physics major research programs are based on the assumption that many phenomena can be described by limiting the number of relevant degrees of freedom to nuclear shape and pairing only. Other examples are the Landau-Ginzburg description of superconductors in terms of a Schr\"odinger-like description for a condensate amplitude, or the effective action formalism in quantum field theory which aims at a description of the dynamics in terms of fewer degrees of freedom. Most of the time it is not  evident {\it a priori} that such an approach is viable, proposed derivations are sometimes misleading, and many approaches rely on intuition and phenomenological arguments. It is particularly important to find examples of physical systems where simplified descriptions are valid and the limits of these descriptions clear.   Equally important are examples that show unsuspected failures of such approaches. We have chosen to investigate pairing dynamics in a fermion system, as this is relevant to a number of different problems: nuclear collective dynamics and nuclear fission in particular, neutron stars, the dynamics of dilute Fermi gases in the unitary regime, quantum hydrodynamics of superfluids in general, effective action description of various strongly interacting Fermi systems from quarks to nuclei to cold gases, and the sometimes invoked relation to a Landau-Ginzburg description of such systems. 

At first we will concentrate our attention on a uniform unitary Fermi gas for a number of reasons: 1) the properties of both homogeneous and inhomogeneous systems are rather well known from {\it ab initio} calculations \cite{carlson,dmc}; 2) the properties of the unitary Fermi gas are very close to the properties of dilute neutron matter, which can be found in the crust of neutron stars \cite{bertsch}; 3) a very accurate description of both homogeneous and inhomogeneous systems is available within a Density Functional Theory (DFT) extended to describe superfluid systems \cite{slda}; 4) the Fermi gas in the unitary regime is under intense experimental scrutiny for a number of years now, see Refs. \cite{ketterle}; 5) a wide spectrum of theoretical approaches of varying sophistication and accuracy have been applied and/or developed for this system, see recent review \cite{giorgini}; 6) the weak coupling limit has been studied extensively within both meanfield approximation and exact treatments \cite{spivak,levitov,others,emil}; 7) the experimental study of the aspects of the pairing dynamics discussed here appears to be feasible in the unitary regime,   see for example a somewhat related experiment \cite{magnetic}. The unitary regime is unlike the weak-coupling regime studied in Refs. \cite{spivak,levitov,others,emil}, where no one has yet suggested a practical experimental realization and verification. Subsequently, we will discuss the case of a non-uniform
unitary Fermi gas, where pairing gap and number density oscillations couple. 

To set the stage, let us review various possible frameworks in which one can study the dynamics of a fermionic superfluid. The oldest approach is quantum hydrodynamics \cite{giorgini}, which can be derived from conservation laws:
\begin{subequations}
\label{eq:qh}
\begin{align}
& \dot{n}  +{\bm \nabla}\cdot[{\bm v} n ]=0,  \\
& m\dot{\bm v} +
{\bm \nabla}\left \{\frac{m{\bm v}^2}{2}+\mu[n ] + V_{ext} \right \}=0,
\end{align} 
\end{subequations}
where  we dropped the arguments $({\bm r},t)$. Above $n({\bm r},t)$ is the number density, 
${\bm v}({\bm r},t)\propto {\bm \nabla} S({\bm r},t)$ is the velocity field determined by the gradient of the phase of the condensate, $\mu[n({\bm r},t)]$ is the chemical potential, and $V_{ext}({\bm r},t)$ is the external field in which the system might reside. One can derive these equations also if one assumes that the magnitude of the condensate remains constant and only its phase $S({\bm r},t)$ can vary. Alternatively, various authors use descriptions based on a Landau-Ginzburg inspired formalisms \cite{babaev,eft}, when the dynamical degrees of freedom are related to the ``condensate wave function"   $\Psi({\bm r},t)$.  In particular these equations describe the Goldstone modes arising from the broken $U(1)$-symmetry in the condensed phase. This wave function is related to the expectation value of the two-fermion field operators 
$\Psi({\bm r}, t)\propto \langle \psi_\uparrow({\bm r}, t)\psi_\downarrow({\bm r}, t)\rangle $, for which one ends up with a Schr\"odinger-like equation
\bea
&& i\hbar\dot{\Psi}({\bm r}, t) = -\frac{\hbar^2{\bm \nabla}^2}{4m}\Psi({\bm r}, t)  \nn \\
&& + U(|\Psi({\bm r}, t)|)\Psi({\bm r}, t) + V_{ext}({\bm r}, t)\Psi({\bm r}, t). \label{eq:lg}
\eea
Here $2m$ is the mass of the Cooper pair and $U(|\Psi({\bm r}, t)|)$ is a Mexican hat-like potential with a minimum at the ground state condensate value $|\Psi_0|$. One can envision two kinds of small amplitude oscillations in these type of models: 1) modes along the valley of the potential, when only the phase of the condensate $\Psi$ changes, which correspond to the expected Goldstone modes of the broken $U(1)$-symmetry; 2) radial Higgs-like excitation modes, when the magnitude of the ``condensate wave function" $\Psi$ varies.  One can derive either of these descriptions as a small amplitude limit of the time-dependent Hartree-Fock-Bogoliubov (HFB) or Bogoliubov-de Gennes (BdG) equations \cite{hfb}. 

The HFB/BdG equations can be regarded also as an approximation to the appropriate time-dependent DFT description of such systems, namely the Time-Dependent Superfluid Local Density Approximation (TD-SLDA), which will be used here. We will show that TD-SLDA allows for a large number of very slow excitation modes, which are absent in either a quantum hydrodynamic, described by Eqs. (\ref{eq:qh}) or a Landau-Ginzburg   description of these systems given by Eq. (\ref{eq:lg}). 

\begin{figure}
\includegraphics[width=7.2cm]{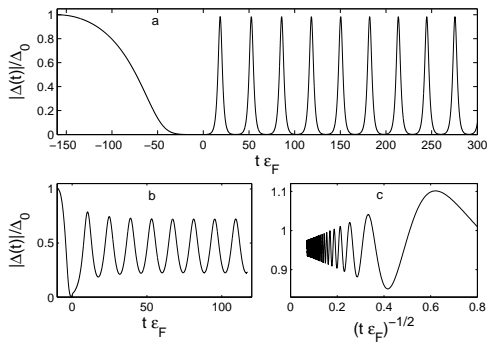}
\caption{ The panels $a, b$ and $c$ display response of the homogeneous system to an initial switching time interval $t_0\varepsilon_F = 160, 10$ and 160 and values of the gap corresponding to $\gamma_s$ are $\gamma_s/\gamma =0.005, 0.005$ and $0.5$ respectively, where $\gamma$ is the coupling constant in Eq. (\ref{eq:ed}) and  $\Delta_0\approx 0.5\varepsilon_F$ is the gap equilibrium value, both at unitarity.   
\label{fig:modes} } 
\end{figure}

The (un-regularized) SLDA  energy density functional has been introduced and discussed in Refs. \cite{slda} (see these references for the description of the renormalization procedure required to eliminate the ultraviolet divergences of the un-renormalized theory and $\hbar=m=1$):
\beq
{\cal E}({\bf r},t)= \alpha \frac{\tau}{2}
                 + \beta \frac{3(3\pi^2)^{2/3}n^{5/3} }{10}
                 + \gamma \frac{|\nu|^2}{n^{1/3}}, \label{eq:ed}\\
\eeq
where $\alpha$, $\beta$ and $\gamma$ are dimensionless parameters and $n({\bf r},t)$, $\tau({\bf r},t)$ and 
$\nu({\bf r},t)$  are the number, kinetic and anomalous densities respectively expressed through the usual Bogoliubov quasi-particle wave function amplitudes $[u_k({\bf r},t), v_k({\bf r},t)]$, with $k$ labeling the quasi-particle states.   
The new element in this Letter is the non-trivial time-dependence of all the quasi-particle wave functions, which formally amounts to replacing the eigenvalues with time-derivatives $E_k\rightarrow i\partial_t$ in the HFB/BdG equations. 
The time-dependent density functional theory is viewed in general as a reformulation of the exact quantum mechanical time evolution of a many-body system when only single-particle properties are considered \cite{gross}. After preparing the system in its ground state, we introduce a time-dependence of $\gamma$ (which controls  the magnitude of the pairing field $\Delta({\bf r},t)$) on a specific schedule. We slowly reduce $\gamma$ in magnitude to a value $\gamma_s$ during a time interval $t_0\gg 1/\varepsilon_F$, after which we rather abruptly bring it back to its value at unitarity in a time interval $\delta t \approx 0.005/\varepsilon_F \ll 1/\varepsilon_F$, see  Fig. \ref{fig:modes} and Refs. \cite{spivak,levitov}, where $n=k_F^3/(3\pi^2)$ and $\varepsilon_F=k_F^2/2$.

This scenario could be realized experimentally by controlling the scattering length with the magnetic field as a function of time, see Ref. \cite{magnetic}. Since by changing the coupling constant alone one does not induce density variations, one might argue that the time-dependence of the meanfield does not a play any role in the dynamics of the pairing field.   At unitarity one has a qualitatively different scenario, since both the meanfield and the pairing field are of the same order of magnitude, and the meanfield $U$ depends strongly on the value of the pairing field, see Refs. \cite{slda},
\beq
U({\bf r},t)= \frac{\beta[3\pi^2n({\bf r},t)]^{2/3}}{2} 
   -\frac{|\Delta({\bf r},t)|^2}{3\gamma n^{2/3}({\bf r},t)}.
\label{eq:pot}
\eeq
We will study a range of phenomena for which the number density $n({\bf r},t)$ is constant in space and time, while $\Delta({\bf r},t)$ will be constant in space only, and this fact alone will lead to changes in $U({\bf r},t)$.  In principle all couplings in Eq. (\ref{eq:ed}) change with the scattering length as in the experiment \cite{magnetic}, but since only $\gamma$ changes drastically from the BCS limit ($k_F|a|\ll 1$) to unitarity ($1/k_Fa=0$), we neglect the changes in $\alpha$ and $\beta$ for now (but reinstate them in a trapped system), which leads only to minor quantitative changes.  One should remember that under these changes of the coupling constant(s) the number density remains constant, and after bringing $\gamma$ back to its value at unitarity the total energy of the system is also conserved. Only during the time intervals $t_0$ and $\delta t$ does one changes the energy of the system. One can excite a large variety of oscillations, and some examples of the collective modes we excite in this manner are shown in Fig. \ref{fig:modes}.  These modes qualitatively resemble those observed in the weak coupling regime 
\cite{spivak,levitov,others,emil}.  

In panels $a$ and $b$ of Fig. \ref{fig:occup} we show the instantaneous single-particle occupation probabilities for the mode shown in panel $a$ of Fig. {\ref{fig:modes}, at times when the pairing field is at its smallest and and its largest values respectively in the oscillatory regime. While the occupation probabilities are essentially identical with their equilibrium values when the pairing gap is almost vanishing, the corresponding distribution at the times when the pairing field reaches its maximum value is drastically different from its equilibrium distribution. Even though the occupation probabilities are so different from their equilibrium values, the corresponding pairing field is hardly different in magnitude from its equilibrium value.  The most surprising feature is the fact that the pairing field oscillates around mean values different from the minimum of the ``effective potential" $ U(|\Psi({\bm r}, t)|)$ in 
Eq. (\ref{eq:lg}), namely $\Delta_0\approx 0.50\varepsilon_F$ \cite{carlson}. One would naively expect that a simpler Landau-Ginzburg-like description as used in Refs.\cite{babaev,eft}, would perhaps be appropriate.  

An interpretation of these modes, even in the small amplitude limit, as being a radial-like oscillation of the pairing field in a Mexican hat-like potential is equally invalid. In the weak coupling limit this was demonstrated by Volkov and Kogan \cite{volkov}, who have shown that the oscillations of the pairing field couple with excited quasiparticles with energies above $2\Delta_\infty$, see also Refs. \cite{levitov,emil}, and that leads for large times to
$  \Delta (t) = \Delta_\infty+A\sin( 2\Delta_\infty t +\phi)/\sqrt{\Delta_\infty t}$. Even though the TD-SLDA equations  have a more complex structure, the dynamics of these modes is very similar in our case, see panel $c$ of Fig. \ref{fig:modes}.
 
The modes displayed in panels $a$ and $b$ of Fig.  \ref{fig:modes} are truly nonlinear and their frequency depends strongly on the oscillation amplitude, see the panels $c$ and $d$ of Fig. \ref{fig:occup}. These are at the same time very slow modes, with frequencies well below the pairing gap $\Omega_H < 2\Delta_0$, but at the same time they are truly large amplitude collective modes, not only because of the size of the oscillation amplitude, but also because their excitation energy is equally large.  In this respect these modes are somewhat similar to the very large amplitude oceanic waves, which take a long time to dissipate into heat. The results of the MIT experiment \cite{magnetic} suggests that the damping of these modes, due to the decay into other modes, is much smaller than one might naively expect at unitarity.  

\begin{figure}
\includegraphics[width=7.2cm]{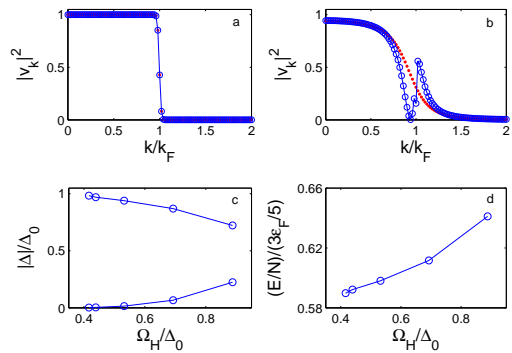}
\caption{(Color online) Panels $a$ and $b$  display the instantaneous occupation probabilities of the mode shown in upper panel of Fig. \ref{fig:modes} corresponding to times $t>0$ when the pairing field is at its minimum and maximum values respectively with circles joined by a solid (blue with circles) line. With (red) dots we plotted the equilibrium occupation probabilities corresponding to the same instantaneous values of the pairing gap. In  panels $c$  and $d$ we show the maximum and minimum values of the oscillating pairing field and the corresponding excitation energy as a function of the frequency of the Higgs-like modes, see Fig. \ref{fig:modes} $a$ and $b$. 
\label{fig:occup}  }
\end{figure}

\begin{figure}[tbp]
\includegraphics[width=8.1cm]{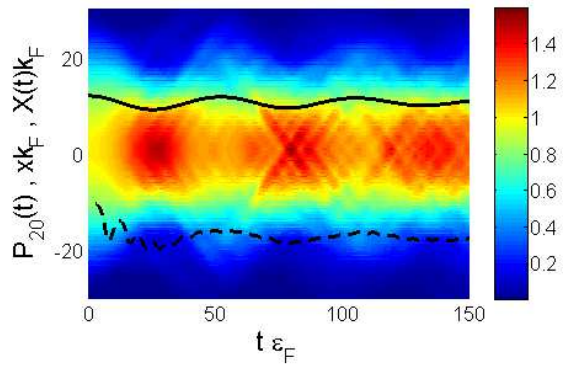}
\caption{(Color online) The color bar shows the correspondence between various values of the ratio   
$n(x,t)/n(0,0)$ and the colors used to represent them. Here $n(0,0)=k_F^3/(3\pi^2)$ and 
$\varepsilon_F=k_F^2/2$. The solid black line shows the corresponding rms cloud radius, see Eq. (\ref{eq:x2}). The dashed black line shows the quadrupole moment of the momentum distribution 
$P_{20}=20 \langle k_y^2+k_z^2-2k_x^2\rangle/(Nk_F^2)$ (scaled to fit in figure).  
\label{fig:cloud}  }
\end{figure}

Since it would be rather difficult to study a unitary homogeneous system experimentally, we have considered the effects one might observe instead in a trapped system. Upon the change of the scattering length the size of a cloud and its central density change and this will induce both number density and pairing gap oscillations. We have considered a semi-realistic case, which we could simulate using our present computer resources, a homogeneous system in $yz$-spatial dimensions ($Lk_F\approx 131$), which is trapped only in the third dimension in a harmonic potential well $V({\bm r}) = \omega^2 x^2/2$ ($\omega/\varepsilon_F\approx 0.0683$ and $N=20,000$ particles). The initial state is that of a very weakly interacting Fermi gas in this potential well and at time $t=0$ we bring quickly the scattering length to its unitary value and keep it constant for the rest of the time evolution of the system. Since the equilibrium radius of a weakly interacting Fermi gas exceeds that of a unitary gas, the system tends to shrink initially and density oscillations of the cloud are thus excited in the $x$-direction. This is unlike the homogeneous case discussed above, where only oscillations of the pairing field were induced. Fig. \ref{fig:cloud} demonstrate a rather complex cloud dynamics. The dynamics of the pairing field is very similar. The rms cloud size in the $x$-direction is described rather accurately 
as simple damped harmonic oscillations
\beq
X(t)=\sqrt{\langle x^2 \rangle}(t) = x_0 + x_1 \exp (-\eta_H t) \cos (\Omega_H t)   
\label{eq:x2}
\eeq
with $\Omega_H/\omega = 1.74$ and $\eta_H/\omega=0.082$ for the case illustrated here and similar ratios for other cases studied by us. The frequency of this mode is consistently lower than that of the quantum hydrodynamic frequency obtained in the small amplitude limit from Eqs. (\ref{eq:qh}) for a unitary gas, namely $\Omega_{QHD}/\omega=4/\sqrt{3}\approx 2.31$.  In this respect this is similar to the behavior of the Higgs-like modes in homogeneous systems. The detailed dynamics is rather complicated and one can identify rather easily running waves, the interference of which leads to ``Landau" damping, 
with $\eta_H\propto v_F/X$ ($X$ is the system Thomas-Fermi radius). This is similar to waves on a surface of a water pool, when multiple reflections from walls lead to a very choppy surface, before the wave energy is converted into heat.  This damping mechanism is different from that discussed in Ref. \cite{turb} (partially already included in the present approach) and likely a more efficient one as well in traps. We estimated the speed of these running waves $v_H/v_{F}$ to be within 10\% of the ground state value of the speed of sound  $c/v_{F}=\sqrt{\xi/3}\approx 0.37$, where $\xi$ is the Bertsch parameter \cite{carlson}. These waves propagate with essentially constant speed, even though the local density or Fermi velocity changes quite dramatically across the cloud. Upon crossing two waves propagating in opposite directions seem to retain their original form (a soliton-like property), in spite of significant nonlinearities. A remarkable feature of this new type of excitation mode of Fermi systems is the intrinsic non-sphericity of the Fermi surface (resembling Landau's zero-sound for normal systems), a  feature absent for the hydrodynamic modes or the collective modes in homogeneous systems described above within TD-SLDA.  Note that in finite systems the equilibrium local momentum distribution is elongated along the density gradient \cite{p2}, thus $\langle k_y^2+k_z^2-2k_x^2\rangle<0$ initially.

In summary, we have discussed the large amplitude pairing field dynamics in both homogeneous and inhomogeneous unitary Fermi systems and have demonstrated the existence collective modes with frequencies lower than the hydrodynamic ones, and which in traps have a non-spherical oscillating  momentum distribution.

The financial support through the US Department of Energy Grants No. DE-FG02-97ER41014 and DE-FC02-07ER41457 is gratefully acknowledged. We thank G.F. Bertsch and M.M. Forbes for their critical reading of the manuscript and discussions, and K.J. Roche for help with implementing the parallel version of the code. Some calculations were performed using the UW Athena cluster.


\end{document}